\documentclass[final,5p,times,twocolumn]{elsarticle}

\usepackage{ecrc}

\usepackage{amsmath}
\usepackage{graphicx}
\usepackage{amssymb}
\usepackage{color,soul}
\usepackage[T1]{fontenc}
\usepackage{bm}
\volume{00}

\firstpage{1}

\runauth{}

\jid{physe}

\usepackage{amssymb}
\begin{document}

\begin{frontmatter}



\dochead{}

\title{An experimental demonstration of the memristor test}


\author[1]{Yuriy V. Pershin\corref{cor1}}
\ead{pershin@physics.sc.edu}

\author[1]{Jinsun~Kim}

\author[1]{Timir~Datta}

\author[2]{Massimiliano Di Ventra}
\ead{diventra@physics.ucsd.edu}

\address[1]{Department of Physics and Astronomy, University of South Carolina, Columbia, South Carolina 29208, USA}
\address[2]{Department of Physics, University of California, San Diego, La Jolla, CA 92093, USA}
\cortext[cor1]{Corresponding author}

\begin{abstract}
A simple and unambiguous test has been recently suggested [J. Phys. D: Applied Physics, {\bf 52}, 01LT01 (2018)] to check {\it experimentally} if a resistor with memory is indeed a memristor,
namely a resistor whose resistance depends {\it only} on the charge that flows
through it, or on the history of the voltage across it. However, although such a test would represent the litmus test for claims about memristors (in the ideal sense), it has yet to be applied widely to actual physical devices.
In this paper, we experimentally apply it to a current-carrying wire interacting with a magnetic core, which was recently
claimed to be a memristor (so-called `$\Phi$ memristor') [J. Appl. Phys. {\bf 125}, 054504 (2019)]. The results of our experiment demonstrate unambiguously that this `$\Phi$ memristor' is {\it not} a memristor: it is simply an inductor with memory. This demonstration  casts further doubts that ideal memristors do actually exist in nature or may be easily created in the lab.

\end{abstract}


\end{frontmatter}

\section{Introduction} \label{sec:1}
The question of whether or not the resistance switching cells are memristors~\cite{chua71a,strukov08a} has intrigued the scientific community since the early days of the modern `memristor era'~\cite{mouttet2012memresistors,meuffels2012fundamental,di2013physical,vongehr2015missing,sundqvist2017memristor,abraham2018case}.
By `memristor' it is meant here one whose resistance depends {\it only} on the the charge that flows
through it, or on the history of the voltage across it~\cite{chua71a}. It is then clear that claims about the existence and nature of these devices have to rely on a {\it direct} and {\it experimental} verification of the functional dependence of the resistance on
either the charge or the flux linkage (time integral of the voltage). However, such a verification has been missing for a long time, thus depriving the scientific community of reliable means to ascertain whether actual physical devices are memristors or not.

Very recently, two of us (YVP and MD) have allayed this issue by introducing a simple memristor test~\cite{pershin18a}.
Moreover, the present authors have applied the test experimentally to Cu-SiO$_2$ electrochemical metallization cells and commercially available electrochemical metallization cells (Knowm, Inc.)~\cite{jkim19a}. Our experimental work~\cite{jkim19a} provides a solid and {\it unambiguous} proof that the resistance switching memories are {\it not} memristors.

However, despite the existence of such an easy experimental test, several claims still plague the scientific literature regarding
the `discovery' of ideal memristors. For instance, very recently, an alleged experimental realization of a `memristor' was proposed by a group of authors~\cite{Wang19a}. In that work, it was suggested that a current-carrying wire interacting with a magnetic core is a `real'  memristor, which the authors called the `$\Phi$ memristor'. It is worth noting that in a comment on that paper written by two of the present authors (YVP and MD)~\cite{pershin2019comment}, the claims of Wang {\it et al.}~\cite{Wang19a} were questioned, and serious concerns were raised about the validity of their results. This resulted in the retraction of Ref.~\cite{Wang19a} from the Journal of Applied Physics based on technical grounds~\cite{Retraction21a}. Such an unfortunate outcome could have been easily avoided if the authors of Ref.~\cite{Wang19a} would avail themselves of the test we had previously suggested. It also shows that such a test has yet to
be fully appreciated by the researchers working in the field.

The purpose of the present paper is then twofold. First, we want to implement the memristor test~\cite{pershin18a} experimentally to
clarify both how the test can
be administered in practice, and, most importantly, its simplicity. Second, we apply it directly to the so-called `$\Phi$ memristor' and
 provide a definitive experimental proof
that the latter is {\it not} a memristor. We thus hope our present paper may serve as a starting point to carefully apply the suggested test, before claims of `discoveries' of such devices are made. Indeed, the present experimental demonstration, is yet another
confirmation that memristors, as defined in Ref.~\cite{chua71a}, may not actually exist in nature, or cannot be built easily in hardware.\footnote{In fact, the closest physical realization of this concept, we are aware of, is the phase-dependent conductance of a Josephson junction, which is, however, just a component of the junction's total current~\cite{Peotta}.}

This paper is organized as follows. In the next Section~\ref{sec-test} we provide the description of the memristor test as formulated in Ref.~\cite{pershin18a}. The goal is to apply it to the `$\Phi$ memristor', which we describe in Sec.~\ref{sec:2a}. In Sec.~\ref{sec:2b}, we introduce the electric circuit used in our experiments. Section~\ref{sec:3a} presents the dynamics of a `$\Phi$ memristor' driven periodically. In Sec.~\ref{sec:3b} we perform the memristor test.  In Sec.~\ref{sec:4} we offer our conclusions.

\section{Methods} \label{sec:2}

\subsection{The memristor test}\label{sec-test}
Let us first introduce the reader the memristor test that was suggested in Ref.~\cite{pershin18a}.
According to it, to prove or disprove that a given device is an ideal memristor or not, the fundamental memristor relation~\cite{chua71a}
\begin{equation}\label{eq:1}
	R=R(q)
\end{equation}
needs to be verified. For this purpose, the tested device is connected in-series with a capacitor, and the circuit is subjected to an arbitrary voltage waveform such that the final capacitor charge equals its initial value (in practice, it is convenient to use the initially discharged capacitor).

According to Eq.~(\ref{eq:1}),  memristors would return to their initial state, each time the capacitor charge returns to the initial value (this is the so-called {\it duality} in the memristor-capacitor circuit~\cite{pershin18a}). In actual experiments,
a sufficiently wide range of voltage amplitudes and initial states of memristors are necessary to prove that the tested device is a memristor. To prove the opposite, a single measurement, in principle, is enough (within the operational range of the device).

%If the supposed memristor does {\it not} return to its initial state after the discharge of the capacitor, then the corresponding device has failed the test: it is {\it not} a memristor.

In the next Section, we apply this test to the `$\Phi$ memristor'. The only distinctive feature of the present implementation of the test is that we will consider the dynamics of magnetization instead of resistance, since the magnetization is the internal state variable in the so-called `$\Phi$ memristor'.

\subsection{The `$\Phi$ memristor'} \label{sec:2a}

To fabricate the `$\Phi$ memristor', one lot of ferrite cores (made by VEB Elektronik Gera, German Democratic Republic in early eighties) was purchased on ebay.  Scanning electron microscopy (SEM) imaging and energy dispersive X-ray (EDS) analysis were carried out using a  TESCAN Vega-3 scanning electron microscope  at the Electron Microscopy Center at the University of South Carolina. Fig.~\ref{fig:1}(a) presents an SEM image of the core. The EDS analysis was performed at three different points on the surface with 4.5 kV accelerating voltage. It was found that the core is composed of ferrite doped by Mg and Mn atoms (about 4 and 8 atomic percent, respectively).

The initial design of the `$\Phi$ memristor'~\cite{Wang19a}  was slightly modified by increasing the coupling between the magnetic core and current-carrying wire, and adding a pick-up coil. For this purpose, two similar coils of three turns each  were wound around the magnetic core using 0.1mm enameled copper wire. The pick-up coil facilitates the reading of the device state: the magnetization reversal events appear as voltage peaks across the pick-up coil. The peak polarity indicates the direction of reversal.

\subsection{Measurement setup} \label{sec:2b}

\begin{figure}[tb]
\begin{center}
(a) \;\;\; \includegraphics[width=50mm]{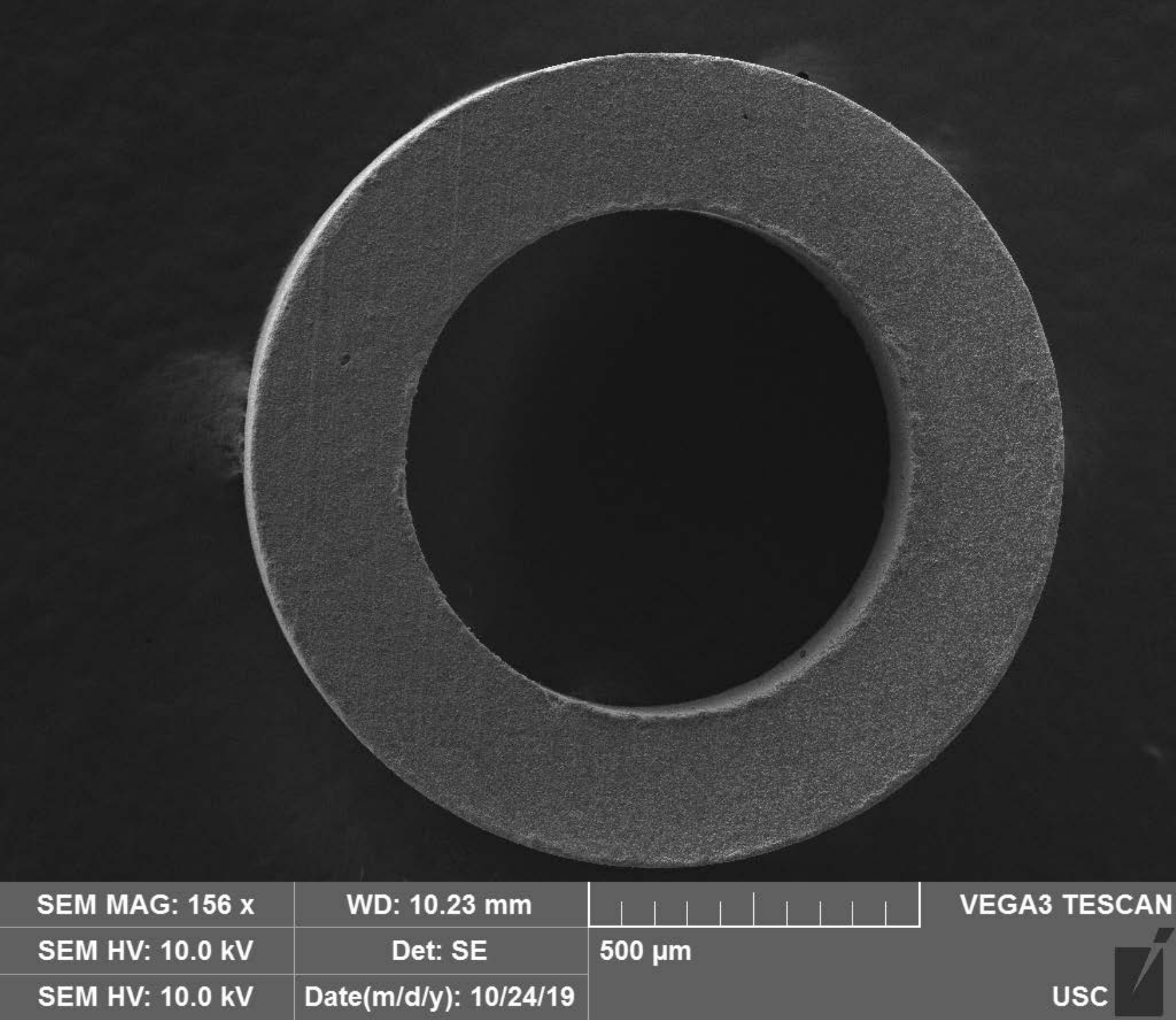} \\
\vspace{0.5cm}
(b) \includegraphics[width=80mm]{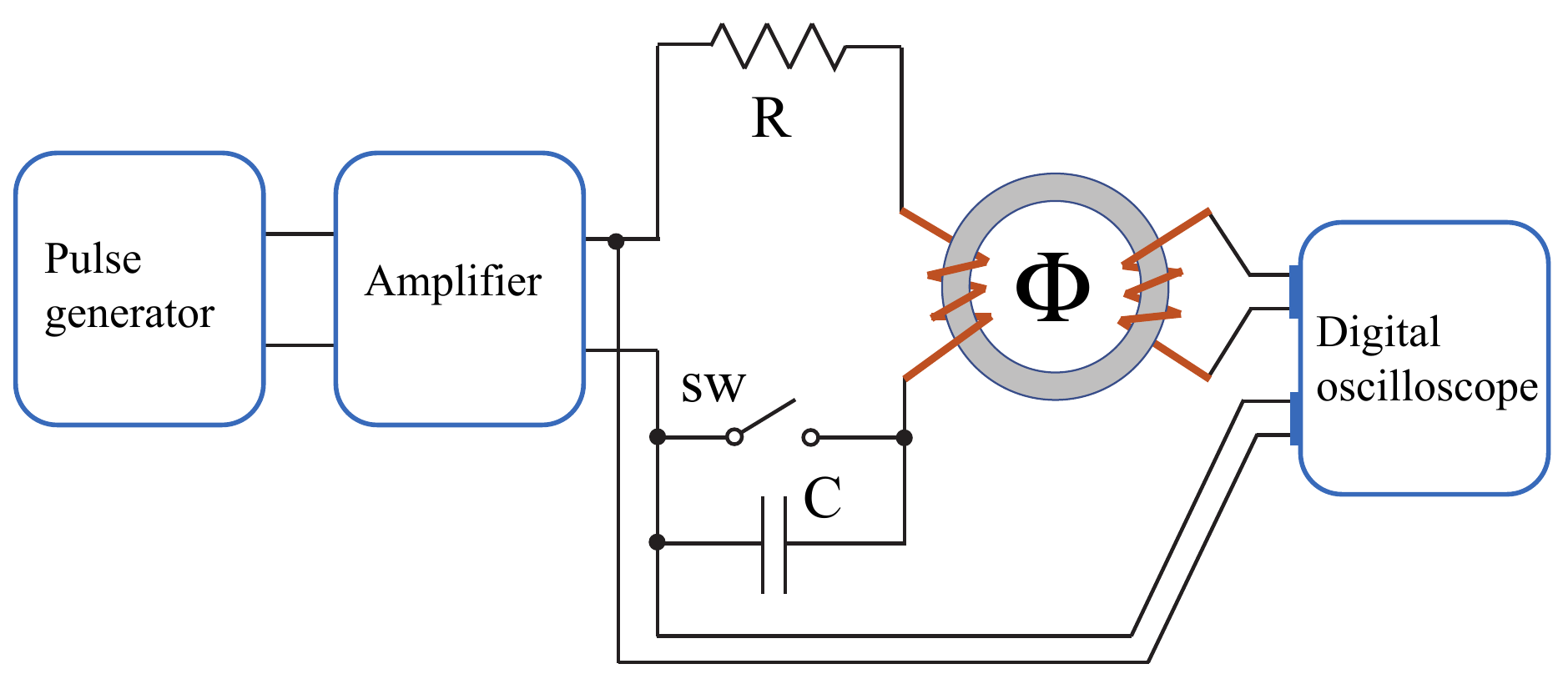}
\end{center}
\caption{(a) SEM image of the magnetic core. (b) Experimental setup for measuring the response of the `$\Phi$ memristor'.}\label{fig:1}
\end{figure}

The basic electrical circuit used in our measurements is shown in Fig.~\ref{fig:1}(b). Here, the standard and arbitrary
waveforms are generated by the arbitrary waveform generator (Siglent SDG1025) and amplified by a power amplifier (HP 467A).
The `$\Phi$ memristor' is connected to the amplifier with the help of a current-limiting resistor (47 $\Omega$) and a
capacitor necessary for the test (0.2 $\mu$F). Two channels of digital oscilloscope (Agilent DSO1012A) measure the voltage across the pick-up coil and applied voltage. To simplify the measurements, the test sequence is applied to the `$\Phi$ memristor' periodically. Some of our measurements were performed at shunted capacitor (closed state of the switch, indicated as SW in Fig.~\ref{fig:1}(b)).

We emphasize that the use of current-limiting resistor does not undermine the test. As the in-series connected resistor does not open any additional charge/discharge channels for the capacitor, the duality between the memristor state and capacitor charge~\cite{pershin18a} is conserved. As an alternative argument we note that the in-series connected resistor and memristor can always be considered as an effective memristor.

\section{Results} \label{sec:3}

\subsection{Periodically driven `$\Phi$ memristor'} \label{sec:3a}

\begin{figure}[tb]
\begin{center}
(a) \includegraphics[width=0.85\columnwidth]{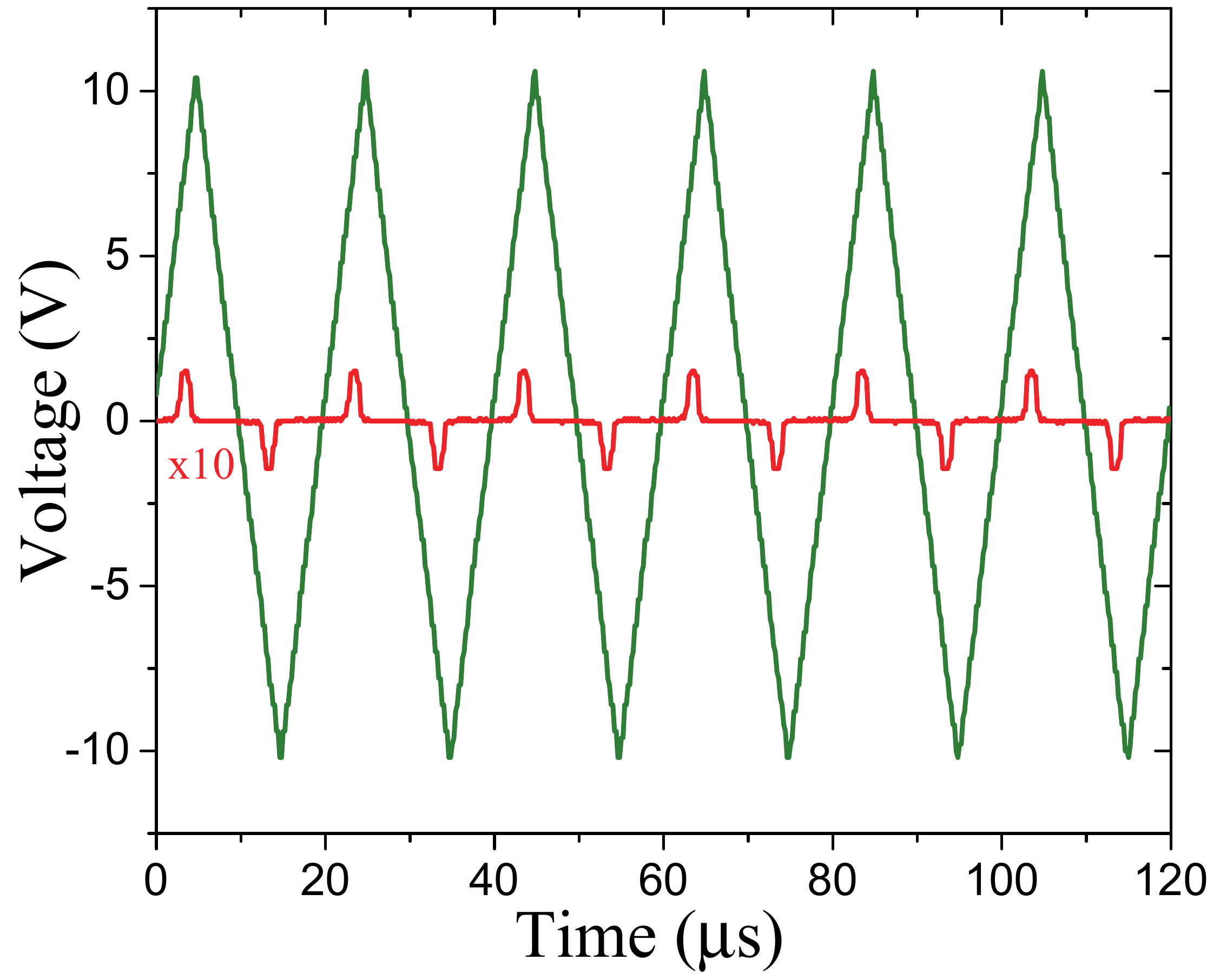} \\
(b) \includegraphics[width=0.85\columnwidth]{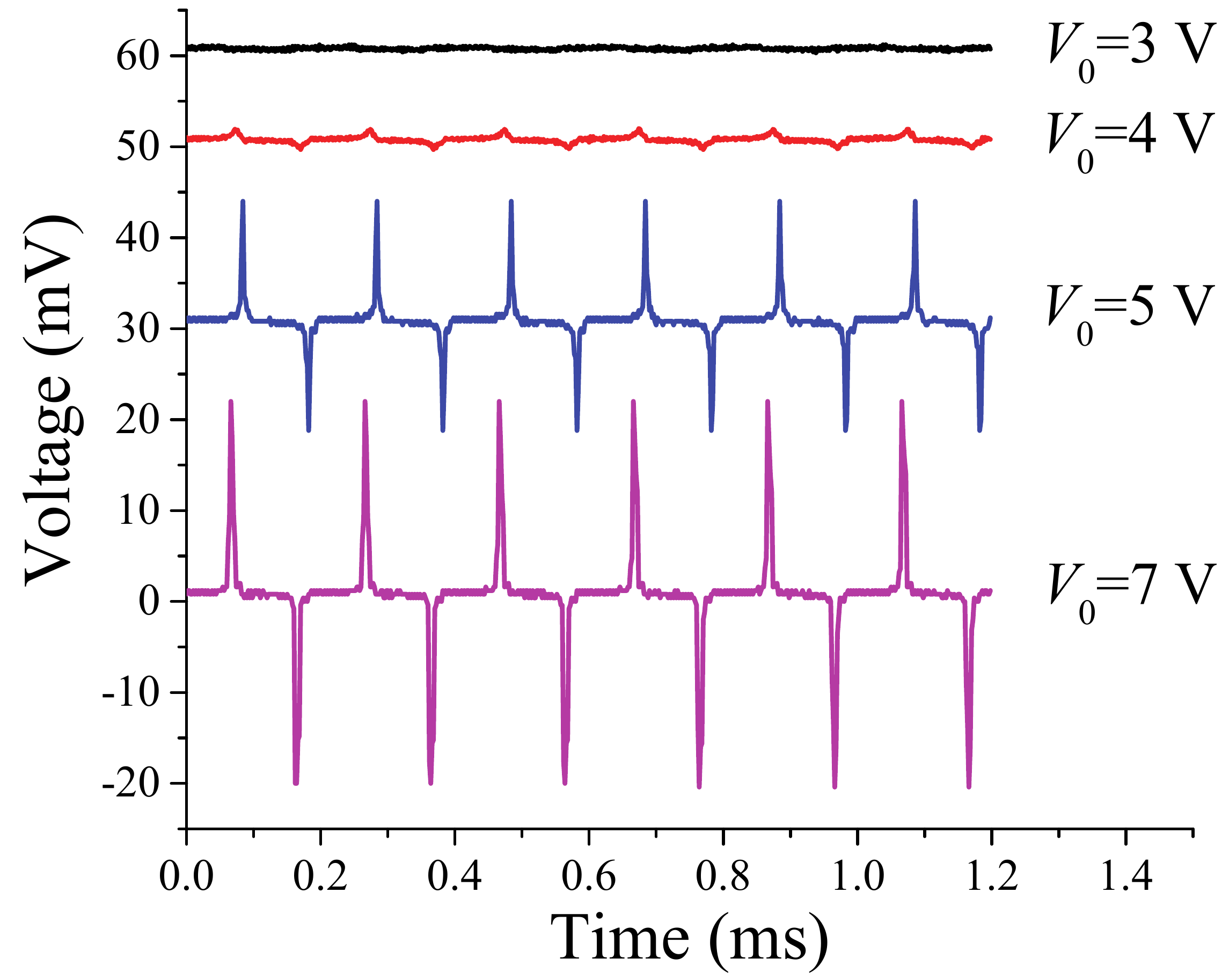}
\end{center}
\caption{(a) Voltage across the pick-up coil (red) for the `$\Phi$ memristor' subjected to a triangular voltage waveform (dark green). To perform this measurement, the capacitor in Fig.~\ref{fig:1}(b) was shunted (by closing the switch SW). (b) Voltage across the pick-up coil for several amplitudes, $V_0$, of a 5~kHz sinusoidal voltage waveform. }\label{fig:2}
\end{figure}

First of all, we verify the basic functionality of the magnetic core and compare results to the literature. For this purpose, we measure the response
of `$\Phi$ memristor' in the circuit of Fig.~\ref{fig:1}(b) with shunted capacitor. The circuit is driven by a sawtooth waveform of $V_0=10$~V amplitude and $f=50$~kHz frequency.  Fig.~\ref{fig:2} shows a clear magnetization reversal pattern, wherein each reversal event appears as a positive or negative voltage peak. We note that such peaks are typical in the response of magnetic core memories (see, for instance, Ref.~\cite{Cooke62a}). Moreover, at smaller voltage amplitudes ($V_0 \lesssim 4$~V) no signal was observed across the pick-up coil. This clearly indicates a hysteretic behavior that can only by related to the magnetic hysteresis of magnetic cores~\cite{Katz69a}.

Our study also involved other periodic waveforms, such as sinusoidal voltages. In all cases, at higher voltage amplitudes ($V_0 \gtrsim 4$~V), and at about the same frequencies, the behavior was qualitatively similar to the one exhibited in Fig.~\ref{fig:2}.

\subsection{Memristor test on the `$\Phi$ memristor'} \label{sec:3b}

Having established that the response of the `$\Phi$ memristor' is typical of magnetic core devices, we are coming now to the question of whether or not it is indeed a memristor. If `$\Phi$ memristor' were a memristor, the core magnetization would be its internal state variable.  Now, a difficulty appears:
at a constant applied voltage, the equilibrium resistance of the `$\Phi$ memristor' does {\it not} depend on the core magnetization, as it is only defined by the wire resistance. This alone already disqualifies the `$\Phi$ memristor' to be a memristor. Nevertheless, we perform the test focusing on the internal state as the resistance of memristors depends on their internal states.

\begin{figure}[t]
\centering{\includegraphics[width=0.9\columnwidth]{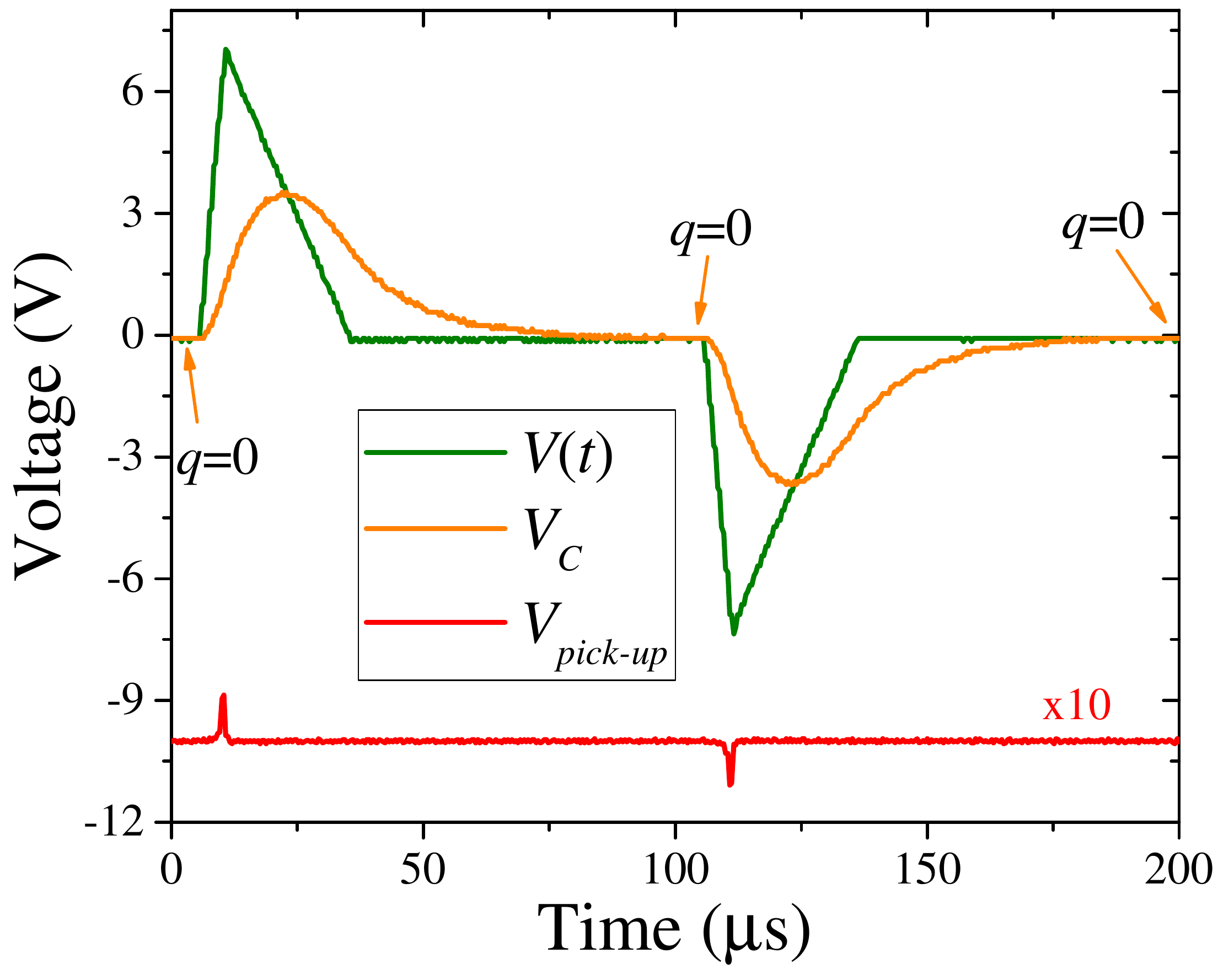}}
\caption{Application of the memristor test to the `$\Phi$ memristor'. The pick-up coil curve (bottom) is shifted for clarity. $V(t)$ is the applied voltage, $V_C$ is the voltage across the capacitor, and $V_{pick-up}$ is the voltage across the pick-up coil. }\label{fig:3}
\end{figure}

The experiment is performed at the open switch SW of Fig.~\ref{fig:1}. The test sequence consists of periodic asymmetric pulses ($V(t)$ curve in Fig.~\ref{fig:3}). In fact, each pulse (either positive or negative) corresponds to a single test. We emphasize that within a half-period of the test sequence (corresponding to a single pulse) the voltage across the capacitor ($V_C$ curve in Fig.~\ref{fig:3}) returns to its initial value (the capacitor is fully discharged right before the beginning of the next pulse). Therefore, all conditions of the memristor test are satisfied~\cite{pershin18a}. If the tested device were a memristor it should then return to its initial state before the next pulse.

However, $V_{pick-up}$ curve in Fig.~\ref{fig:3} shows a dramatically different behavior. The magnetization flips to one direction by a positive pulse, and reverses only by the application of the negative one. In other words, the memristor-capacitor duality property we
have mentioned in Sec.~\ref{sec-test} is not satisfied. Thefore, we conclude that the `$\Phi$ memristor' fails the memristor test, and  {\it cannot} be described by Eq.~(\ref{eq:1}).

We repeated this experiment with several other peak amplitudes, and the `$\Phi$ memristor' response was always the same. Therefore, our conclusion is that the `$\Phi$ memristor' is not a memristor.

\section{Conclusion} \label{sec:4}

Although it should have been evident from the start that a current-carrying wire interacting with a magnetic core is simply an inductor, we have nonetheless applied the memristor test suggested in Ref.~\cite{pershin18a} to the `$\Phi$ memristor' as an example and to disprove directly the claims in Ref.~\cite{Wang19a}. The test unambiguously demonstrates that such a device is not a memristor. Also, it is important to notice that the deviations of the `$\Phi$ memristor' behavior from the memristor model in Eq.~(\ref{eq:1}) are {\it too significant} to be described by
small modifications of such an equation. This supports the memristor impossibility conjectures formulated in Ref.~\cite{jkim19a}, regarding the difficulty of building a model of physical resistance-switching memories based on the memristor model.

Finally, while in this work we have focused on a specific device, the
test is general. Indeed, it should be applied to all those devices that are declared as 'memristors', to check the validity of such claims.

\section*{Acknowledgment}

This research was partially supported by an ASPIRE grant from the University of South Carolina.

\bibliographystyle{elsarticle-num}
\bibliography{memcapacitor}

\end{document}